\pdfoutput=1

\documentclass[%
 preprint,
 amsmath,amssymb,
 aps,
pra,
]{revtex4-2}

\usepackage{graphicx}
\usepackage{dcolumn}
\usepackage{bm}
\usepackage{setspace}

\begin{document}

\preprint{Pre-print Copy}

\title{Elimination of Extreme Boundary Scattering via Polymer Thermal Bridging in Silica Nanoparticle Packings: Implications for Thermal Management}
\thanks{This work is currently under review.}%

\author{Brian F. Donovan}
\thanks{These authors contributed equally to this work.}
\email{bdonovan@usna.edu}
\affiliation{Department of Physics, United States Naval Academy, Annapolis, MD, 21402, USA}
\author{Ronald J. Warzoha}
\thanks{These authors contributed equally to this work.}
\affiliation{Department of Mechanical Engineering, United States Naval Academy, Annapolis, MD, 21402, USA}
\author{R. Bharath Venkatesh}
\affiliation{Department of Chemical and Biomolecular Engineering, University of Pennsylvania, Philadelphia, PA, 19104, USA}
\author{Nicholas T. Vu}
\affiliation{Department of Mechanical Engineering, United States Naval Academy, Annapolis, MD, 21402, USA}
\author{Jay Wallen}
\affiliation{Department of Mechanical Engineering, United States Naval Academy, Annapolis, MD, 21402, USA}
\author{Daeyeon Lee}
\affiliation{Department of Chemical and Biomolecular Engineering, University of Pennsylvania, Philadelphia, PA, 19104, USA}
\email{daeyeon@seas.upenn.edu}


\date{\today}
\setstretch{1.5}

\begin{abstract}
{\bf Recent advances in our understanding of thermal transport in nanocrystalline systems are responsible for the integration of new technologies into advanced energy systems, including thermoelectric refrigeration systems and renewable energy platforms. However, there is little understanding of heat energy transport mechanisms that govern the thermal properties of disordered nanocomposites. In this work, we explore thermal transport mechanisms in disordered packings of amorphous nanoparticles with and without a polymer filling the interstices in order to quantify the impact of thermal boundary scattering introduced at nanoparticle edges in an already amorphous system and within the context of a minimum thermal conductivity approximation. By fitting a modified minimum thermal conductivity model to temperature-dependent measurements of thermal conductivity from 80 K to 300 K, we find that the interstitial polymer {\it eliminates} boundary scattering in the disordered nanoparticle packing, which surprisingly leads to an {\it increase} in the overall thermal conductivity of the disordered nanoparticle thin-film composite. This is contrary to our expectations relative to effective medium theory and our understanding of a minimum thermal conductivity limit. Instead, we find that a stiff interstitial material improves the transmission of heat through a nanoparticle boundary, improving the thermal properties of disordered nanoparticle packing. We expect these results to provide insight into the tunability of thermal properties in disordered solids that exhibit already low thermal conductivities through the use of nanostructuring and vibrational thermal bridging.}
\end{abstract}

\maketitle


The effective medium description of thermal transport in composite materials expresses thermal conduction by a weighted average of its individual constituents \cite{warzoha2014heat, minnich2007modified}. This approach can be extremely useful to describe thermal transport in systems ranging from simple bulk composites to complex thin films with nanoparticle inclusions \cite{kumar2007effect, warzoha2014effect, zheng2011thermal}. The conventional effective medium approach, however, utilizes the intrinsic thermal conductivity of each constituent and does not account for potential changes in thermal transport mechanisms upon combination or nanostructuring of each material. Without modification, thermal conductivity as described by the effective medium approach is therefore limited by the highest constituent thermal conductivity in the composite. Modifications to the effective medium approach with respect to nanoparticle inclusions have been made to account for thermal boundary conductance between the inclusions and the matrix \cite{minnich2007modified}; however, these modifications do not permit larger composite thermal conductivity than either of the individual components. In this work, we utilize a material system that consists of a disordered nanoparticle packing with and without interstitial polystyrene to show that extreme levels of boundary scattering can be alleviated by establishing a thermal bridge between nanoparticles and their surroundings. This thermal bridging mechanism enhances the composite thermal conductivity of our material system beyond the thermal conductivity predicted by effective medium approximations that make use of each individual constituent's thermal conductivity. Consequently, these results demonstrate the potential to tune thermal properties of disordered solids through additional nanostructuring in tandem with a new ability to regulate heat energy carrier boundary scattering.

Beyond their use to demonstrate this remarkable thermal interaction, disordered nanoparticle packings represent a critically important class of materials that have profound implications for optical components \cite{hakim2007nanoparticle}, surface coatings \cite{xia2004directed}, water desalination materials \cite{niksefat2014effect} and electronic applications \cite{lin2017enhanced}. However, less is known about how heat moves through such disordered solids. Without knowledge of the mechanisms that drive thermal transport in such material systems, their successful integration into device platforms is unlikely to be realized. For instance, disordered nanoparticle thin-films have recently been proposed for solar-driven water desalination \cite{chan2014tailoring}. In this specific application, the thermal conductivity of the thin-film structure is critical to achieving sufficient performance for complete desalination. To this end, we investigate the mechanisms that drive thermal transport across the thin-film structure. 

Recent developments in our understanding of thermal transport in disordered materials have shown that despite accurate modeling of a mean free path that is described by the minimum limit to thermal conductivity \cite{cahill1992lower}, thermal conduction in these materials can be further reduced due to intrinsic boundary scattering \cite{hopkins2011ultra, goodson2007ordering, pernot2010precise}. This is particularly important to consider for nanoparticle thin-films, where it is likely that both the film boundary and the individual nanoparticle boundaries contribute significantly to thermal transport through these systems. To account for scattering across fine particle boundaries, of which there are many in disordered nanoparticle packings, the nanoparticle boundaries are treated similarly to grain boundaries within analytical models to understand the underlying mechanisms that govern thermal transport across the entire film \cite{donovan2014spectral}. A widely accepted means to account for this additional phonon scattering mechanism comes from the introduction of an additional mean free path limitation that is proportional to the diameter of the grain or nanoparticle, which itself is directly related to the thermal conductance across the individual nanoparticle boundary (or boundaries) \cite{wang2011thermal}. Significant work has also been done to achieve thermal bridging and promote heat flow across boundaries by matching materials and vibrational properties \cite{szwejkowski2017molecular, maire2017heat, ravichandran2014crossover}. In this work, we show that the voids between nanoparticles lead to an extremely large boundary scattering mechanism and that by the infiltration of polystyrene into the interstitial voids, we are able to eliminate the effects of that boundary scattering and bridge between the nanoparticles and polymer as one effective composite film. 

In order to understand the physical mechanisms that regulate heat flow in this amorphous composite material system, we model the thermal conductivity of both the polystyrene as well as the silica nanoparticles using a Debye model that is modified to account for scattering as described by the minimum limit to thermal conductivity \cite{cahill1992lower}. Using this model, we are able to account for some spectral dependence of the thermal conductivity, but avoid the incorporation of a full density of propagon, diffuson, and locon states for both constituents in this system, which would otherwise be highly computationally intensive. Instead, we employ an effective medium approach to capture the dynamics of both the nanoparticle packing and the infiltrated polymer \cite{minnich2007modified}. We find that this simple model can be applied to our complex material system to elicit the physics that govern thermal transport in each material system, and is the only model that captures a large increase in the thermal conductivity of the composite effective medium compared to either constituent. Additionally, this effective medium approach coupled with a complete elimination of the nanoparticle boundary scattering allows us to capture the temperature dependence of the thermal conductivity of the composite film (from 80 K to 300 K) without any fitting parameters. This finding implies that the incorporation of an additional amorphous material filling the interstitial voids in disordered nanoparticle packings relaxes the extreme boundary scattering and returns the system to one that is dominated by scattering due to intrinsic amorphous thermal carriers. 






Capillary rise infiltration (CaRI)\cite{huang2015polymer, hor2017nanoporous, jiang2018toughening} was used to make the polymer-infiltrated composite films. Silicon wafers (N-type, 500 $\mu$m thicknees, 100 mm circular wafers, 0-100 $\Omega\cdot$cm electrical resistance), used as substrates, were purchased from University Wafers Inc. The wafers were cut into 1 x 1 cm$^2$ sizes, cleaned with isopropanol, acetone, and water, and then plasma cleaned for 4 minutes. A commercial aqueous suspension of Ludox TM-50 silica nanoparticles was purchased from Sigma Aldrich. The suspension was diluted to 15 wt$\%$, sonicated in an ultrasonic bath for 4-5 hours, and then filtered using a 0.45 micron hydrophilic syringe filter purchased from Fisher Scientific. The colloidal solutions are spin-coated on top of the cleaned silicon wafers at 2200-2500 RPM for 1.5 minutes. Multiple coatings were used to make layers of thickness 600-650 nm without any cracks \cite{prosser2012avoiding}. 

Polystyrene (PS) of molecular weight 173000 g/mol (PS-173k) was purchased from Polymer Source Inc. Solutions of 3$\%$ and 2$\%$ PS-173k in toluene were made and then filtered using 0.20 microns hydrophobic PTFE syringe filter purchased from Sigma Aldrich. PS-173k solutions were spin-coated atop the cleaned silicon wafers at 2500-3000 RPM to make films of thickness 100-110 nm and 160-170 nm from the 2$\%$ and 3$\%$ solutions, respectively.

\begin{figure}[h!]
  \includegraphics[width=16cm]{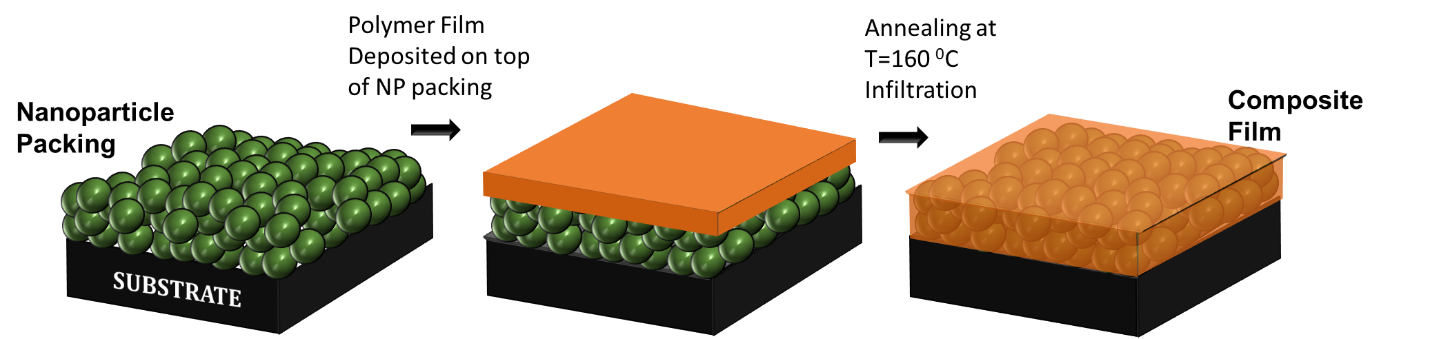}
  \caption{Schematic illustrating the capillary rise infiltration (CaRI) method. A polymer film is collected on top of a spin-coated nanoparticle film and then annealed at a high temperature to make a CaRI composite film.}
  \label{infiltration}
\end{figure}

A polymer film on the silicon substrate was cut at the edges using a razor blade without damaging the substrate. The film was then immersed slowly into a water bath at an angle of 45$^{\circ}$ to the water surface such that the PS film delaminates from the substrate and floats on the water surface. The disordered nanoparticle film, spin coated earlier on the silicon substrate, was then immersed in the water bath to capture the floating film on top of the film. The bilayer was then dried overnight and annealed at 165$^{\circ}$C for 12 hours to get complete infiltration of the polymer into the voids of the nanoparticle packing to get the CaRI composite film. The ratio of the polymer film thickness to the nanoparticle packing thickness was chosen such that the pores were filled upon full infiltration. A schematic of the polymer infiltration procedure is provided in Fig. \ref{infiltration}.

We note that the silica nanoparticles are hydrophilic and are cast in the form of a disordered packing from an aqueous solution by using spin-coating. They contain hydrophilic hydroxyl groups on the surface. When a thin polymer film in contact with the nanoparticle packing is annealed, the capillary forces are able to induce imbibition of the hydrophobic polymer into the pores of the nanoparticle packing thus bringing the two phases in intimate contact with each other. Because of the high surface energy of silica nanoparticles, PS has a fairly low contact angle which facilitates the capillary rise infiltration process.

The volume fraction of nanoparticle was determined by using the refractive index obtained from ellipsometry and expressing it as a simple volume-mixing mean of the two components that make the nanoparticle film – air and nanoparticles. So, the refractive index n is written as,

\begin{equation}
   n = n_{silica}\phi_{silica}+n_{air}(1-\phi_{silica})
\end{equation}

\noindent assuming that the amount of capillary condensed water in the packing is very small. Taking the values of the refractive index of pure silica and air to be 1.45 and 1 at 632.8 nm wavelength, we solve for $\phi_{silica}$ which is equal to 65\%. We would also like to point out that this result is in keeping with the volume fraction of spheres in a close-packed, random, disordered arrangement of spheres where the void space has been reported to be around 34\% \cite{remond2008characterization}.The void fraction in the packing can also be determined by the amount of polymer that completely fills the interstices of the packing upon infiltration. This is estimated by calculating the difference between the thickness of the polymer film before and after infiltration – $h_o$ and h respectively. The polymer occupying the voids after complete infiltration is given by,

\begin{equation}
    V_{polymer}=\phi_{void} h_{NP} A
\end{equation}

\noindent where A is the base area of the packing $h_{NP}$ is the height of the nanoparticle packing. $V_{polymer}$ can also be expressed as the amount of polymer that is consumed from the polymer film, which is given by,

\begin{equation}
   V_{polymer} = A(h_o - h)
\end{equation}

\noindent Equating the two, we can express the porosity as,

\begin{equation}
    \phi_{void} = \frac{(h_{o}-h)}{h_{NP}}
\end{equation}

The volume fraction of the nanoparticles in the packing is 65\% and stays constant throughout. The free surface of the silica nanoparticles inside the packing is entirely covered with polymers in the case of the completely-filled CaRI films.

We also note that when the UCaRI films (or the undersaturated CaRI films) are synthesized, the polymer spreads out into the particle packing and is mostly located at the necks between adjacent nanoparticles. As the fill fraction of polymer increases, it spreads out into the bulk of the packing and covers all free surfaces inside the packing.

The thermal conductivity of each thin-film stack was measured using the optical pump-probe technique Frequency-domain Thermoreflectance (FDTR). FDTR utilizes a thin ($\sim$ 80 nm) metal transducer at the top of the sample stack to allow for heat absorption from one continuous wave ``pump'' laser. A separate continuous wave ``probe'' laser is used to track the reflection of the sample. The reflectivity of the transducer is known to change with temperature, and a phase lag is generated at the sample surface at a particular frequency due to a corresponding lag in the temperature response at the sample surface upon heating \cite{schmidt2009frequency}. To maintain a high sensitivity to the temperature response at the surface, the wavelength of the probe beam is centered at the wavelength corresponding to the transducer's maximum coefficient of thermoreflectance. In this case, we utilize a 532 nm probe wavelength to capture reflectivity as a function of applied pump frequency. Consequently, by sweeping through a wide range of frequencies one can track the frequency-dependent phase lag, or surface temperature, of the sample stack. In this case, we utilize a 405 nm pump laser (Coherent OBIS 405 LX) to establish the heating event on the sample surface and a separate 532 nm probe laser (Coherent OBIS 532 LS) to measure the frequency-dependent response of reflectivity at the surface of the transducer by monitoring voltage via a balanced photodiode. This measured response is used to fit a multilayer thermal model that, in this experiment, provides sensitivity to the thermal conductivity of the films primarily, but also can be used to verify the heat capacity of the film. We note that in fitting the model to the data, we also leave the unknown transducer-film interfacial conductance as a free parameter, though our sensitivity analysis indicates that it does not affect our fitted values significantly. Additional details pertinent to this specific system can be found in the recent study by \citet{sharar2019solid} and the Experimental Details section at the end of the manuscript. 

We use an analytical model for the thermal conductivity based on a Debye assumption to capture the thermo-physical mechanisms that drive heat flow within the CaRI composite film made of disordered silica nanoparticle packing and polystyrene \cite{kittel1976introduction, kittel1949interpretation}. Since both constituents are comprised of amorphous materials, we turn to this simplified model to elicit a physical construct that describes thermal transport without making an undue number of assumptions or relying on the computational intensity of {\it ab initio} simulations. In our model, we determine the thermal conductivity, $\kappa$ by

\begin{equation}
    \kappa = \int_0^{\omega_D}\hbar\omega \mathrm{DOS}(\omega)\frac{\partial F(\omega)}{\partial T}v(\omega)^2\tau(\omega) d\omega
\end{equation}

\noindent where we integrate over vibrational frequency, $\omega$. DOS represents the Debye density of states, $F$ is the Bose-Einstein distribution, $v$ is the longitudinal speed of sound in the material \cite{proctor1971sound,mott2008sound}, and $\tau$ is the vibrational scattering time. The cutoff frequency used, $\omega_D$, is determined by the sound speed and the edge of the Brillouin zone given typically by $\pi/a$. Since we are dealing with amorphous materials, we determine the value for $a$ by taking a cube root of the literature values for the molecular density \cite{kittel1949interpretation}. 

To properly model heat energy carrier scattering in amorphous nano-scale systems, we assume that the dominant scattering mechanisms are boundary scattering (both film and nanoparticle length scales for the disordered silica nanoparticle packing) and intrinsic scattering, which is represented by the minimum limit to thermal conductivity \cite{cahill1992lower}. These scattering mechanisms are taken into account by Mathiesenn's rule.

\begin{equation}
    \frac{1}{\tau}=\frac{1}{\tau_{min}}+\frac{1}{\tau_{film}}+\frac{1}{\tau_{NP}}
\end{equation}

\noindent We treat $\tau_{film}$ using a conventional approximation and divide the film thickness by the sound speed \cite{sellan2010cross}, though we ultimately find this to contribute negligibly to the total scattering within the material system. The intrinsic scattering time, represented by the minimum limit to thermal conductivity, is known to capture the thermal transport in bulk amorphous materials well \cite{cahill1987thermal}, and is expressed as $\tau_{min}=\omega/\pi$.

The nanoparticle boundary scattering term is expected to have a significant impact on the thermal conductivity of the disordered nanoparticle packing. Since the nanoparticles themselves are packed, hard spheres, there is a significant fraction of each particle that is thermally isolated from each other (surrounding) particle. As a result, we expect the typical formulation for boundary scattering to be insufficient in describing the inability for thermal carriers to conduct across nanoparticle boundaries. Consequently, we turn to a boundary scattering term described by Wang \textit{et al.}, where an additional factor related to thermal transmission across the boundary, $\alpha$ is used to capture an increased reduction in thermal conductance \cite{wang2011thermal}. This nanoparticle boundary scattering term is embodied by the following expression.

\begin{equation}
    \tau_{NP}=\frac{\alpha d}{v}
\end{equation}

\noindent where $d$ is the nanoparticle diameter. 

In the CaRI composite film, we use the effective medium approach, where the total thermal conductivity, $\kappa_{tot}$, is modeled by a weighted average of the constituent thermal conductivities as follows \cite{minnich2007modified}

\begin{equation}
    \kappa_{tot}=V_{poly}\kappa_{poly}+V_{NP}\kappa_{NP}
    \label{effmed}
\end{equation}

\noindent where $V_{poly}$ and $V_{NP}$ are the volume fractions of the polymer and nanoparticles, respectively. In this study the respective volume fractions are $V_{poly} = 0.35$ and $V_{NP}=0.65$ for the disordered nanoparticle packing that is infiltrated with polystyrene \cite{huang2015polymer}. We note that for the case of the disordered silica film (without infiltrated polystyrene), we consider $V_{NP}$ = 0.65 and $V_{void}$ = 0.35 (the volume of the voids) to compare effective medium approximations between different films using an overall volume. Here, the thermal conductivity of the voids is set equal to 0 due to the lack of any material between the nanoparticles under vacuum. 

We use this functional form of the thermal conductivity to compare to the measured data of cross-plane (i.e. out-of-plane) thermal conductivity measurements from 80 K - 300 K in order to better understand the physical mechanisms that govern thermal transport in nano-engineered amorphous materials. It is important to note that the only parameter in this model that is not defined by either literature values or the dimensions of our system is the boundary scattering transmission factor, $\alpha$.

\begin{figure}[h!]
  \includegraphics[width=16cm]{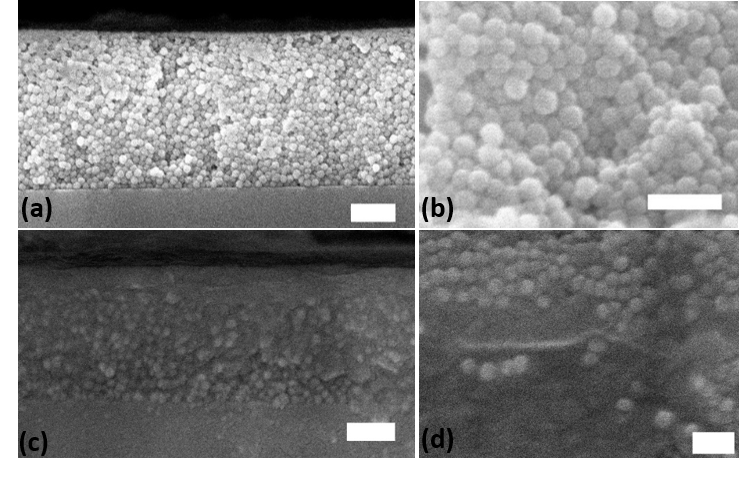}
  \caption{(a and b) Cross-sectional SEM images of the nanoparticle packings before infiltration. (c and d) CaRI composite film after infiltration. Scale bars in (a and c) = 200 nm; scale bars in (b and d) = 100 nm.}
  \label{SEM}
\end{figure}

The structure of the CaRI composite films was characterized using scanning electron microscopy (SEM) by observing the cross-section images of the neat nanoparticle packings without polymer (before infiltration) and nanoparticle packings with polymer in the interstices (after infiltration).  The cross-section images show that the void spaces in these highly disordered nanoparticle packings is completely filled by the polymer after CaRI. The void spaces between the particles, which appear lighter and unfilled in Fig. \ref{SEM}(a) are filled by the infiltrated polymers in the composite films in Fig. \ref{SEM}(c). This is also evident in the edges of the nanoparticles losing the smooth roundness in Fig. \ref{SEM}(d) which is visible in Fig. \ref{SEM}(b).

\begin{figure}[h!]
  \includegraphics[width=16cm]{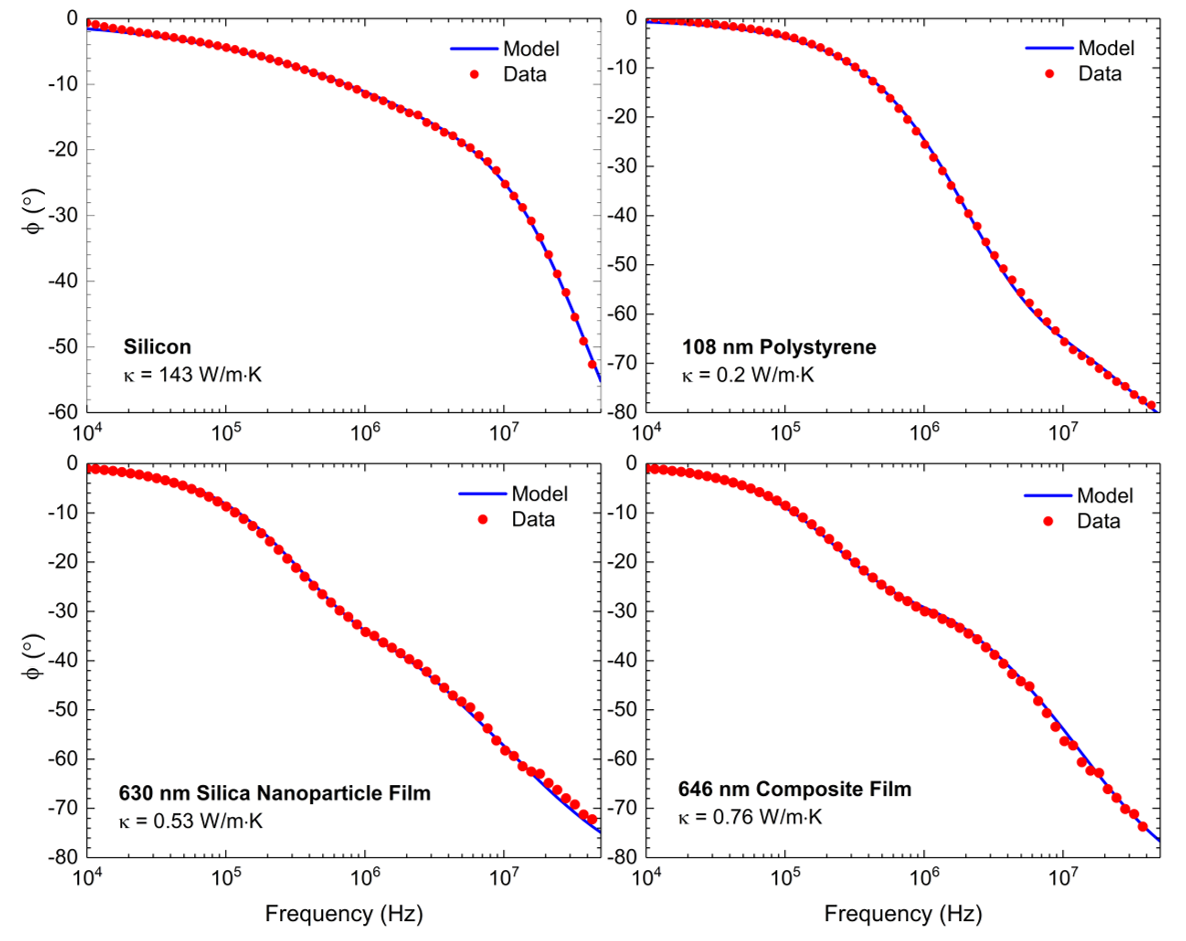}
  \caption{Frequency-phase plots for Silicon (upper left), 108 nm polystyrene on Si (upper right), 630 nm silica nanoparticle film (lower left) and 646 nm composite film (lower right). Solid red circles represent the measured frequency at each phase and blue solid lines represent a fit of the multi-layer analytical model used to extract $\kappa$.} \cite{schmidt2009frequency}
  \label{phase}
\end{figure}

The measured phase lag for each sample using FDTR is provided in Fig. \ref{phase} for the case when all samples are at 300 K. We perform a sensitivity analysis\cite{schmidt2009frequency} to ensure that we are sensitive to both $\kappa$ and C$_v$, which are left as free parameters in this work. We measure temperature-dependent values for C$_v$ and find that they are within 8\% of literature data for each of the constituent materials in the system. We leave the thermal boundary conductance as a free variable in this analysis but do not report on the exact fitted values as we are relatively insensitive to it in our fitting procedure.


Interestingly, the thermal conductivity for the silica nanoparticle film ($\kappa$ = 0.53 W/m$\cdot$K) is significantly reduced from literature values for bulk SiO$_2$ thermal conductivity, which is $\sim$ 1.2 W/m$\cdot$K at 300 K \cite{cahill1987thermal}, even if the packing fraction of silica nanoparticles ($\sim$ 0.65) is accounted for. This indicates that there is additional scattering beyond the proposed minimum limit to thermal conductivity and, despite being amorphous and having an assumed near-atomic mean free path, boundary scattering from the nano-structuring and thermal isolation of the nanoparticles is substantially limiting vibrational transport. 

\begin{figure}
    \centering
    \includegraphics[width=15cm]{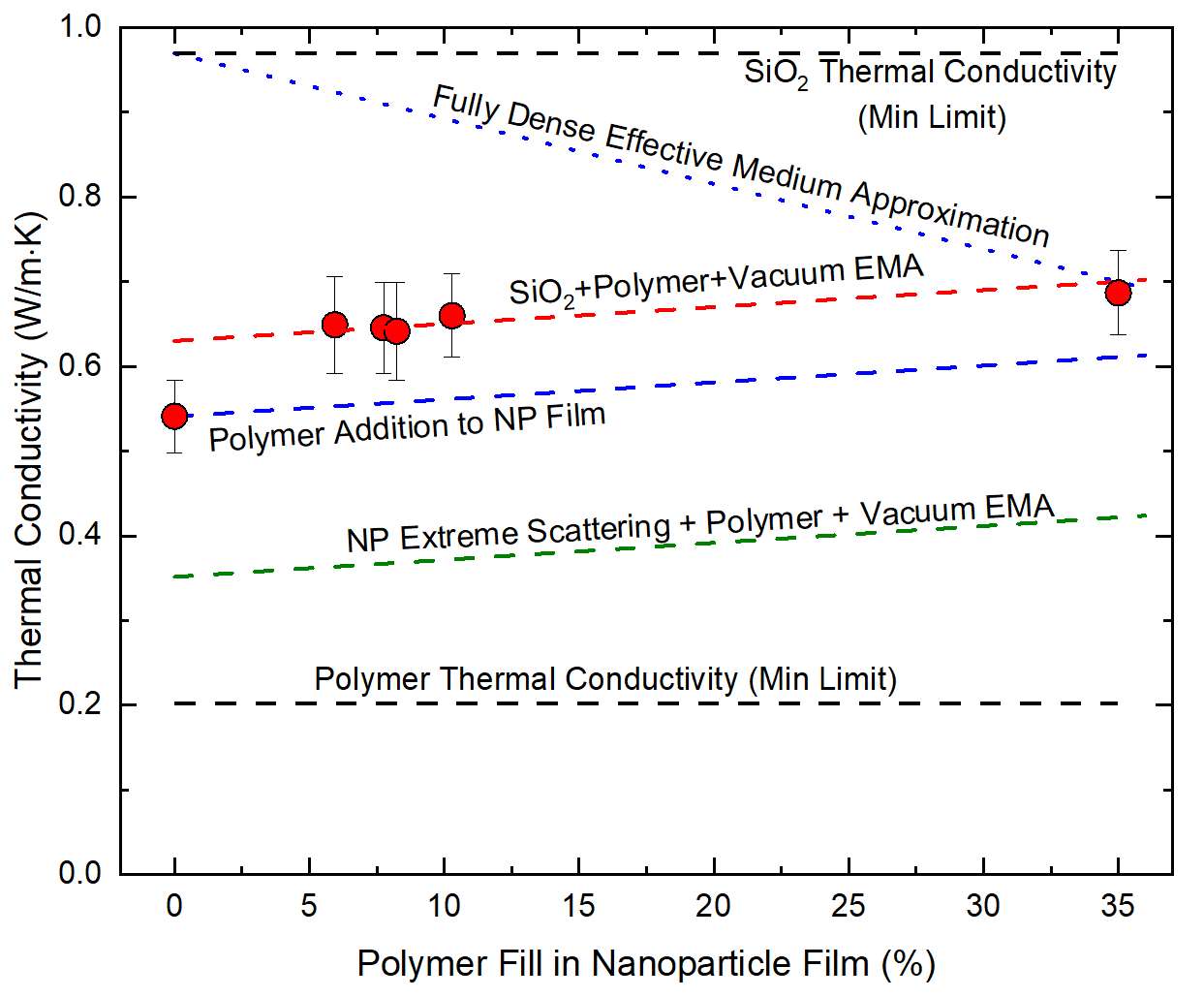}
    \caption{Measured thermal conductivity with varying polymer fill percentages (maximum fill is 35\%). Also included are the thermal conductivity of the polymer and SiO$_2$ calculated with the minimum limit model (black dashed lines). Additionally, we have included the effective medium approximation (EMA)  for the fully dense combination of the constituents (blue dotted line), EMA for a three component system where the polymer addition replaces vacuum in a three component system but the nanoparticle scattering is relaxed to the SiO$_2$ contribution (red dashed line), a direct addition of the polymer to the measured SiO$_2$ nanoparticle film (blue dashed line), and the EMA for the polymer replacing vacuum but maintaining the extreme scattering from the nanoparticle boundaries (gren dashed line).} 
    \label{kappavfill}
\end{figure}

Perhaps more interesting is that the measured thermal conductivity of the composite film is greater than the thermal conductivity of either constituent, even if we treat the problem as if the polymer increases the available volume for heat transfer (dotted green line discussed later in Fig. \ref{kappavfill}, which relates $\kappa_{eff}$ = 1.0$\cdot$$\kappa_{NP}$ + 0.35$\cdot$$\kappa_{polymer}$). Because the thermal conductivity of the polymer is instrinsically low, it stands to reason that by introducing the polymer into the disordered nanoparticle packing, we have relaxed the mechanism that is most responsible for reducing the thermal conductivity of the silica constituent of the composite when in nanoparticle film form (relative to that of bulk SiO$_2$). 

\begin{figure}[h!]
  \includegraphics[width=15cm]{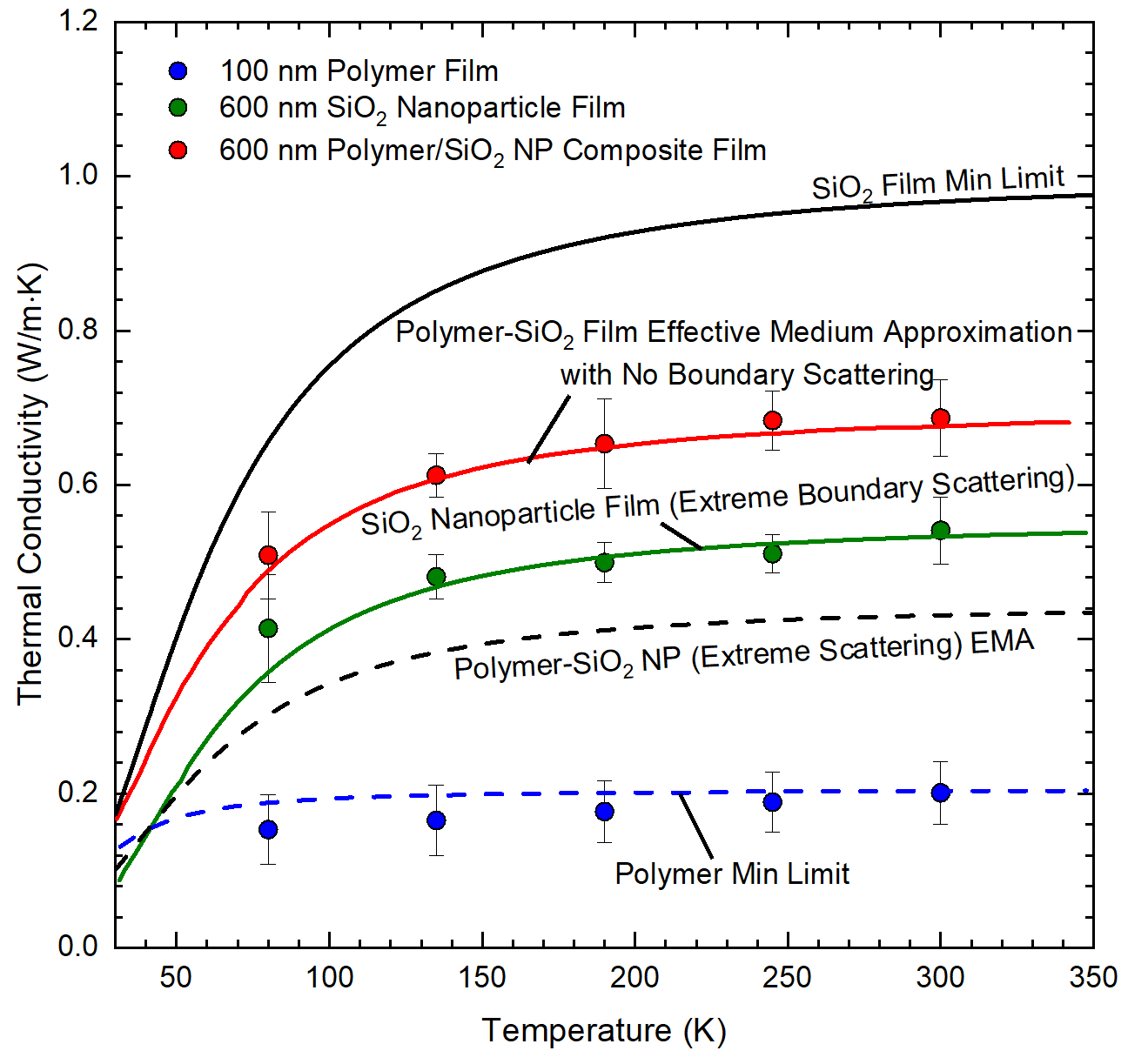}
  \caption{Measured thermal conductivity of the 100 nm polymer film (blue circles), 600 nm SiO$_2$ nanoparticle (NP) films (green circles) and 600 nm SiO$_2$ NP/polymer composite films from temperatures ranging from 80 - 300 K. Also included is the minimum limit to thermal conductivity model for the polymer (blue dashed line) as well as an SiO$_2$ thin film (black solid line). These models are used to understand the trends in the data as the nanoparticle packed bed exhibits extreme boundary scattering (green solid line), and the composite data is only captured by an effective medium comprised of SiO$_2$ and the polymer but without the extreme nanoparticle boundary scattering (red solid line). The effective medium with boundary scattering does not capture the data and is also featured (black dashed line).}
  \label{model}
\end{figure}

To further understand these effects and build intuition about the magnitude of this change in thermo-physical properties, we turn to our thermal model and compare our results to measurements taken over a wide temperature range. As seen in Fig. \ref{model}, we have taken measurements of each sample (polymer film, silica nanoparticle bed, and composite film) at temperatures that range between 80 - 300 K. Accompanying this data is a set of models as described previously, all based on the minimum limit to thermal conductivity with additional scattering from the nanoscale structure of the packing, that are used to gain physical insight into thermal transport across each material. The model for the thermal conductivity of the disordered silica nanoparticle packing utilized a nanoparticle scattering proportionality that is quite small ($\alpha = 0.03$) in order to capture the magnitude of the reduction compared to the bulk thermal conductivity of silica, whereas the thin-film polymer abides by the minimum thermal conductivity limit across its temperature range.

In addition to the minimum thermal conductivity models for the individual constituents, we have included the composite models based on the effective medium approach (Eqn. \ref{effmed}). One of these curves shows the effective medium approach using the reduced thermal conductivity of the silica nanoparticle film in tandem with the thermal conductivity of the polymer at each temperature. This curve clearly does not capture the data and would not be expected to as we see that the composite thermal conductivity is actually higher than that of each individual constituent. 

If we focus on the model used to fit for the silica nanoparticle thermal conductivity and use it in the effective medium calculation approach with the additional scattering from the nanoparticle boundaries removed (i.e. removing the only fitted parameter in this model, $\alpha$) we find that we capture the composite film data without any fitted parameters and the system can be described by an effective medium of the polymer and SiO$_2$ with no nanoparticle scattering.

Physically, these results indicate that the polymer in the interstices of the particle packing effectively relaxes the scattering of vibrational thermal carriers at the nanoparticle interfaces through vibrational bridging. This has significant implications not only in the understanding of the increased composite thermal conductivity, but of the nature of the severe reduction in the thermal conductivity of the disordered nanoparticle packing to begin with; this implies that the severe boundary scattering ($\alpha<<1$) arises specifically from the thermal isolation of the nanoparticles when they are simply packed in. If one examines the morphology of both the silica nanoparticle film and the composite film, it becomes clear that the packing of these hard spheres leads to a relatively small fraction of physical contact with other spheres over which thermal conduction can take place. The remaining surface area of each nanoparticle is exposed to vacuum for these measurements (we note that the frequencies used for the FDTR measurement correspond to thermal transport times that are sensitive to conduction heat transfer mechanisms only, and not convection or radiation \cite{hopkins2012minimum}). Due to this thermal isolation, we see that the minimum limit, or molecular mean free path of vibrational carriers, does not wholly capture the thermal conductivity of the nano-structured material (despite the typical interpretation of the minimum limit as it applies to amorphous systems, which implies that carriers scatter so readily that extrinsic effects such as boundary scattering are not readily realized). 

The addition of the polymer in the interstices, however, seems to reinstate the typical minimum limit interpretation of this material system. The effective medium approach that maintains the boundary scattering will not capture the increase in composite thermal conductivity. However, the elimination of boundary scattering in the nanoparticle constituent of the effective medium (i.e. modeling the intrinsic thermal properties of the amorphous silica without boundary scattering), actually captures the data without any fitted parameters. This excellent agreement is not altogether surprising as there have been a number of studies that have shown that the minimum limit to thermal conductivity actually captures the thermal conductivity of amorphous and disordered materials well \cite{cahill1987thermal, cahill1992lower}. However, this switching off of the scattering of the thermal carriers due to the presence (or lack thereof) of isolated boundaries is quite remarkable. With the infiltrated polymer as a vibrational bridge, the low intrinsic mean free path thermal carriers in the silica nanoparticles are able to transmit through the film and we revert to the physical system where these carriers are no longer significantly affected by the boundaries. This also implies that a potentially low thermal boundary conductance at the contacting regions between nanoparticles is not the principle mechanism for reductions in the thermal conductivity of the disordered silica nanoparticle film.

The relaxation of extreme boundary scattering by vibrational bridging is further emphasized via measurement of nanoparticle films that are only partially intercallated with polymer. In Fig. \ref{kappavfill}, we show measurements of nanoparticle films with varying polymer fill percentages. In order to understand the results of this measurement, we have included a variety of different models to track the addition of the polymer into the system. The first set of conclusions to be made from this set of measurements is that simple effective medium approximations, that is, a fully dense effective medium of SiO$_2$ and the polymer and the effective medium between the nanoparticle scattered film and the polymer filling interstices (replacing vacuum), do not capture the data at all. As mentioned previously, a simple addition of the polymer thermal conductivity to the thermal conductivity of the nanoparticle film does not capture the data once the film is being filled.

We see that once we start adding any polymer to the packed nanoparticle film, we immediately see an increase in the thermal conductivity of the composite system regardless of amount of polymer. Previous studies of these material systems have shown that any addition of polymer (from low percentages, to the complete fill at 35\%) result in the polymer completely coating the nanoparticles \cite{hor2017nanoporous}. This means that with any addition of polymer, we would immediately gain the relaxation of the extreme nanoparticle boundary scattering as the polymer coating provides the vibrational bridge between particles. This fits precisely with model of an effective medium of an SiO$_2$ film with no boundary scattering and a polymer intercallation that replaces the vacuum in the interstices. This model, shown by the red dotted line in Fig. \ref{kappavfill} captures this data set as well without any fitting parameters. This remarkable correspondence of our model to the data from films that have any polymer included, and the lack of correspondence to the other possible interpretations of the system, further highlights the understanding that this jump in the thermal conductivity of the composite system is due to vibrational bridging relaxing the extreme nanoparticle scattering.

We note that the models used in this study rely on a number of simplifying assumptions and thus may not capture the complete temperature trend in this data. However, we use these simplified models to guide our physical understanding of the principle mechanisms that govern the observed changes in thermal transport in the nanoparticle thin-films studied here. With respect to the ``minimum limit'' model, the Debye model for thermal conductivity is not inclusive of certain complexities associated with the actual thermal transport physics in these materials. For instance, the linear dispersion used will lead to an overestimation of high frequency vibrations carrying heat in the polymer and silica nanoparticles. This could be a major reason for our over estimations of the thermal conductivity at low temperatures, where a real dispersion would exhibit thermal transport behavior favored by long wavelength vibrations and would be more affected by boundaries. Further, the ``minimum limit" model clearly does not capture the entirety of the thermal physics that regulate thermal conductivity, even in the case of the pure polymer film. In this case, we ignore the amorphous nature of the vibrations in favor of a simple phonon-like scattering factor. If we were to take into account the dynamics of propagon transport separate from diffuson and locon transport, we may find that the models used for the individual nanoparticle film and the polymer film (i.e. the minimum limit models) fit the data better, particularly at low temperatures. Despite these limitations, the physical mechanisms outlined by the models used in this work do well to describe the differences in thermal conduction between the composite film and each of the constituent films and are not expected to result in alterations to our physical interpretations of the thermal transport physics within the films provided that we achieve any improvement in the fit due to the inclusion of propagon physics via {\it ab initio} simulations. As a result, we maintain that these models provide a deep level of physical insight with zero fitting parameters (or only one in the case of the nanoparticle film).

We have demonstrated a material system where a reduction in the thermal conductivity beyond the well-known ``minimum limit" is achieved by thermo-physically isolating the nanoparticles at boundary regions that are not already in contact with other nanoparticles, which is attributed to extreme boundary scattering at these surfaces. Furthermore, we have demonstrated that this severe reduction can be mitigated and the thermal conductivity can be significantly increased through the introduction of a vibrational bridge between the nanoparticles via capillary rise infiltration (CaRI). Owing to the fact that all of these constituents are amorphous, we are able to capture the data for a CaRI composite thin-film using a simple Debye model for thermal conductivity and a minimal mean free path with the elimination of the extreme boundary scattering. This result can significantly advance our understanding of nanocomposite thermal conductivity due to the fact that we are able to isolate the thermal transport in the disordered nanoparticle film without the presence of the surrounding polymer matrix, which has not been measured and modeled in such a way that these effects can be separated within the wider nanocomposite literature. 

Beyond the thermo-physical implications, this study is expected to provide context and information for the engineering of disordered nanoparticle packings in a variety of novel applications, including as optical coatings, solar-thermal desalination materials, electronics thermal management materials and thermal insulation materials. In particular, this highlights that even with amorphous nanoparticles, the use of nanostructuring (in this case disordered nanoparticle packing) can substantially reduce the thermal properties of amorphous materials beyond what can be predicted using the minimum thermal conductivity limit. Within the context of this work, we expect that further exploration of this boundary scattering modification and ultimately {\it tuning} by modifying parameters such as the stiffness of the polymer (chemically, or by moving through the glass transition temperature), which is likely to correspond to a range of designed thermal conductivities that can be tailored to a particular application.

\section{Acknowledgements}

BFD and RJW thank Mr. Peter Morrison and the Office of Naval Research for their financial support under Contract No. N0001419WX00501. RJW also thanks Dr. Mark Spector and the Office of Naval Research for their financial support under Contract No. N0001419WX00312. RBV and DL thank Penn MRSEC (DMR 1720530) for support, and Syung Hun Han (University of Pennsylvania) for his assistance with sample preparation.

\newpage
\section{Experimental Details}
{\bf Film Thickness Measurements.} The thickness of the bare nanoparticle packings, polymer films, and the CaRI composite films were all measured using an Alpha-SE spectroscopic ellipsometer from J.A. Wollam using a wavelength range of 370-900 nm. The source incidence and detection angles are set at 70$^{\circ}$ relative to normal incidence. The Cauchy model, which relates the refractive index to the wavelength as n = A + $\frac{B}{\lambda^{2}}$ + $\frac{C}{\lambda^{4}}$, was used to fit the raw data from the ellipsometer to extract thickness values. The bare nanoparticle packings and the composite films measured 600-650 nm and the polymer films measured 100-110 nm. The Mean Square Error (MSE), reported by the CompleteEASE Software after fitting the data to the model, was kept as low as possible to get the most accurate value of thickness. 

{\bf Scanning Electron Microscopy.} Scanning electron microscopy (SEM) images of the top-down and cross-section of the samples (without the gold layer) were taken using a JEOL 7500F HRSEM. Cross-section images were taken by cleaving the sample using a diamond scribe and mounting the sample vertically on a stub with the cross sections facing the electron beam. The samples were sputtered with a 4 nm iridium layer using a Quorom plasma generating sputter coater prior to imaging to prevent charging. An accelerating voltage of 5 kV, emission current 20 $\mu$A, at a working distance of $\sim$8 mm was used to image the samples.

{\bf Frequency-Domain Thermoreflectance Measurements.} As described in the manuscript, the optical pump-probe technique Frequency-Domain Thermoreflectance (FDTR) is used to measure the thermal properties of the samples studied in this work. The samples used in this study are coated with an $\sim$ 80 nm Au transducer due to its high coefficient of thermoreflectance at 532 nm. To deposit a gold layer atop the CaRI composite films, the samples were attached on a 10 cm diameter clean silicon wafer using double-sided tape and placed into the E-beam thermal evaporator (PRO Line PVD 75, Kurt J. Lesker Company, PA, USA) in an inverted manner such that the sample faces downward. With a deposition rate of 0.1 A/s, gold was thermally evaporated and deposited on the substrate evenly. At desired thickness of 80 nm, the deposition was terminated and sample was retrieved from the evaporator. The Au film thickness was confirmed with profilometry for each sample and varied at most by $\pm$ 7.2 nm. This was used in the calculation of uncertainty of our thermal conductivity measurements.

Frequencies between 10 kHz and 50 MHz are applied to the pump beam by amplitude modulation of the laser source through an electro-optic modulator (EOM, KDP M350, Model 200 Driver), which is driven directly by our lock-in amplifier (Zurich UHFLI). We lock into the modulated response of the probe beam as we sweep through the frequencies applied to the pump beam in order to determine phase at the sample surface. The modulated pump beam phase is recorded by measuring the phase as a function of frequency directly via the balanced photodetector, permitting electronic noise cancellation by signal substraction of the pump and probe beams. Finally, a multi-layer analytical model is applied to extract relevant thermal properties ($\kappa$, G and C$_v$).

The thermal conductivity of each sample is measured as a function of temperature by placing each sample into a Montana Instruments Cryostation 2. The Cryostation housing includes an x-y-z translation stage and an Agile Temperature Sample Mount (ATSM), which allows for rapid changes in sample temperature as well as precision control of sample temperature (uncertainty U$_T$ $\approx$ 0.025 K). We pump and probe through a quartz window outfitted above the sample, whereby we lose $\sim$ 2 $\%$ of the intensity deposited onto the sample surface and reflected back through the window off of the sample surface. We also note that the steady-state (DC) temperature rise \cite{braun2018steady} is kept below 1$^{\circ}$C due to the presence of the Si substrate beneath the sample stack.

\newpage
\bibliography{achemsodemo}

\end{document}